\documentclass[twocolumn]{aastex631}
\usepackage{bm}
\usepackage{graphicx} 
\usepackage{float}
\usepackage{amsmath}
\usepackage{parskip}
\usepackage{orcidlink}

\begin{document}

\title{Quadratic frequency dispersion in the oscillations of intermediate-mass stars}

\author{Subrata Kumar Panda}
\affiliation{Department of Astronomy and Astrophysics, Tata Institute of Fundamental Research, Colaba, Mumbai 400005, Maharashtra, India}

\author{Shravan Hanasoge \orcidlink{0000-0003-2896-1471}}
\affiliation{Department of Astronomy and Astrophysics, Tata Institute of Fundamental Research, Colaba, Mumbai 400005, Maharashtra, India}

\author{Siddharth Dhanpal \orcidlink{0000-0001-8699-3952}}
\affiliation{Department of Astronomy and Astrophysics, Tata Institute of Fundamental Research, Colaba, Mumbai 400005, Maharashtra, India}

\author{Vageesh D. C. \orcidlink{0009-0002-4206-5390}}
\affiliation{Microsoft Research India, Bengaluru 560001, Karnataka, India}

\begin{abstract}
    Asteroseismology, the study of stellar vibration, has met with great success, shedding light on stellar interior structure, rotation, and magnetism. Prominently known as $\delta$ Scutis, intermediate-mass main-sequence oscillators that often exhibit rapid rotation and possess complex internal stratification, are important targets of asteroseismic study. $\delta$ Scuti pulsations are driven by the $\kappa$ (opacity) mechanism, resulting in a set of acoustic modes that can be challenging to interpret. Here, we apply machine learning to identify new patterns in the pulsation frequencies of $\delta$ Scuti stars, discovering resonances spaced according to quadratic functions of integer mode indices. This unusual connection between mode frequencies and indices suggests that rotational influence may play an important role in determining the frequencies of these acoustic oscillations.
\end{abstract}

\section{Introduction}

Serving as a bridge between the low and high-mass stars, intermediate mass ($1.5-2.5M_\odot$) main sequence stars, which fuse core Hydrogen for a prolonged time, constitute an important group of objects in our Galaxy. Ages, rotation rates and chemical compositions of these stars remain largely uncertain, which when constrained, can substantially help in the {study of Galactic evolution}. {These stars oscillate across a broad spectrum of frequencies, and provided a suitable theory of oscillations, may be used to discern their interior structure, rotation and composition.} Commonly known as $\delta$ Scutis, these stars are notable for their rapid rotation (\citealt{altair}) and associated oblateness (\citealt{altairScience}).

Concurring well with theoretical expectations (\citealt{Lignieres, Reese2006, Reese2009, Reese2017}), photometric observations from space telescopes including MOST, CoRoT, Kepler, K2 and the ongoing TESS have unraveled a class of $\delta$ Scutis that oscillate with uniformly spaced acoustic modes (\citealt{Matthews, HD174936, reg-MOST, reg-corot, TOUCAN, Nature2020}) in a way the solar-like stars pulsate. Modes in $\delta$ Scutis are driven by the $\kappa$ mechanism (\citealt{kp_mechanism}), i.e., opacity fluctuations in He ionization zones in the envelopes induce acoustic oscillations. Consequently, mode amplitudes are not set in accordance energy equipartition, as is the case with turbulently driven oscillations in solar-like stars.

Among the 1000 $\delta$ Scuti stars observed in the sectors 1-9 of TESS, only 60 stars oscillate with regularly spaced modes (\citealt{Nature2020}). Extending this work by using TESS observations till sector 63, 6711 more $\delta$ Scutis were identified (Singh et al. 2024, in review), among which 436 stars show regular pulsation. These two analyses emphasize that some 6\% stars among the $\delta$ Scuti population vibrate in acoustic modes equally spaced in frequency. {This small fraction is representative of the reduced likelihood of capturing stars in the near-ZAMS (Zero-Age Main Sequence) phase of evolution (during which they exhibit periodically spaced oscillations, \citealt{Nature2020}) owing to them spending only a small fraction of their extended main-sequence lifetimes in this state.}

These patterns are characterized by the large frequency separation ($\Delta\nu$) - the inverse diametric sound-crossing time - which is directly related to mean stellar density. In general, $\Delta\nu$ is a function of frequency and thus shows departure from uniform spacing \citep[e.g.,][]{Nature2020}. \cite{HD174936} have quantified the non-linear frequency dependence of $\Delta\nu$ for a $\delta$ Scuti star observed by CoRoT. This characteristic is a well known feature of pulsating stars and typically arises from structure considerations and not necessarily due to rotation.

These 6\% of stars have modes that may be labelled, thereby allowing for interpreting their origin and inferring the stellar structures, constraining stellar ages, open clusters, and moving groups (\citealt{Nature2020}). The remaining 94\% of the population represent stars in main-sequence proper, are rapidly rotating, with substantial structural deformation, providing a wealth of physical insight into the evolution of intermediate-mass stars. However, the oscillations of these stars do not obviously show regularly spaced oscillatory modes, making the interpretation of their spectra challenging. Here, we propose to explore whether at least some of these stars show unusual or unexpected dispersion relations. Making progress thus requires discovery of new and theoretically unanticipated structure in observed oscillation spectra. Provided a new pattern is established in a group of stars, characterizing the mode pattern may point to related time scales and dynamics. Here, we developed a machine-learning based methodology to discover patterns in oscillation spectra, which upon application to observations identified a new resonant-mode pattern. This success paves the way to a more ambitious discovery-oriented examination of $\delta$ Scutis spectra.

\section{Result}

\begin{figure*}
    \centering
    \includegraphics[width=\textwidth]{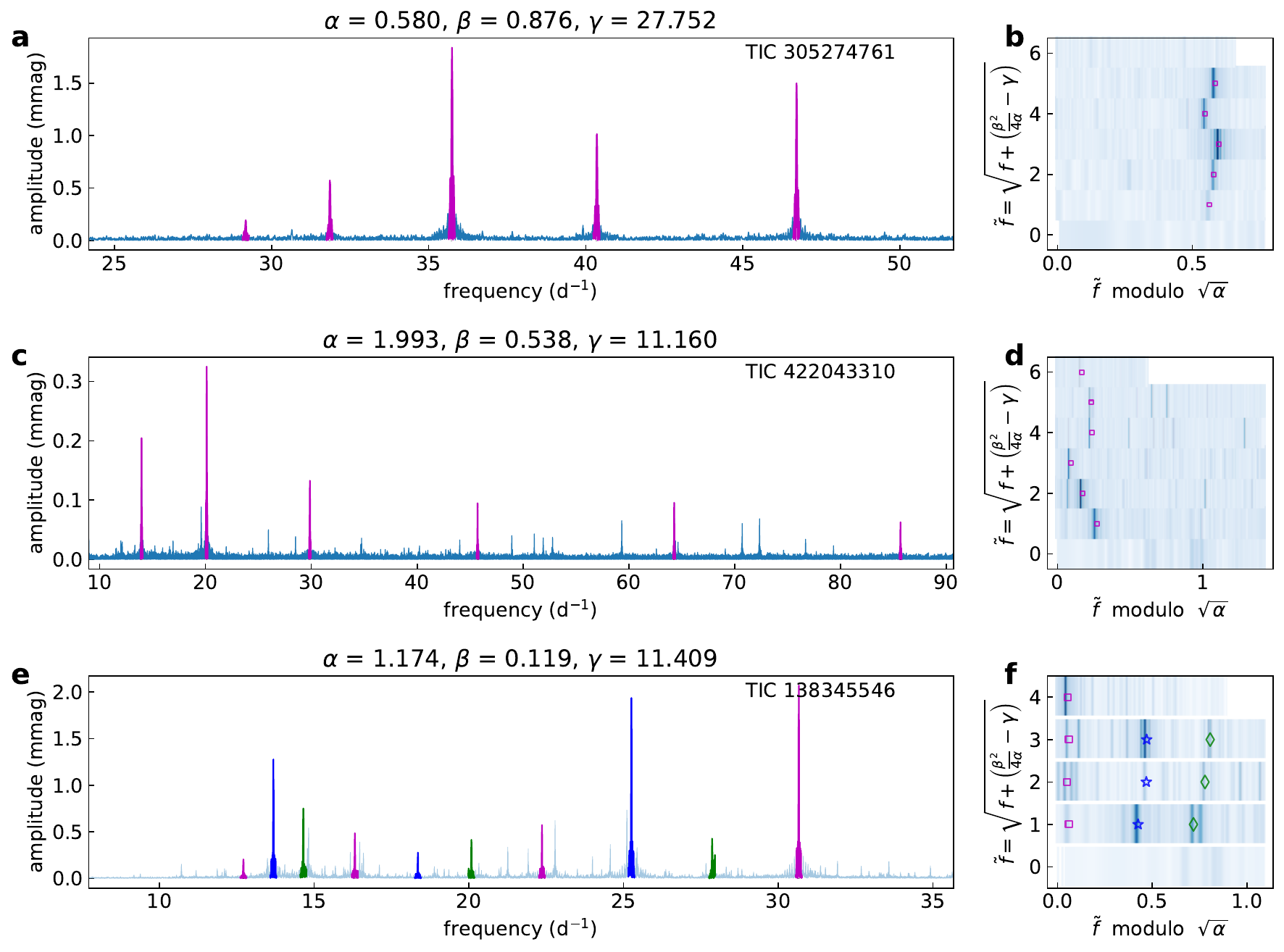}
    \caption{{\bf Quadratic oscillation patterns observed in TESS spectra of $\delta$ Scutis.}
    {\bf a.} Oscillation spectra of TIC 305274761, which pulsates in only five modes. The spacing between adjacent modes keeps increasing, and the frequencies are well described as a quadratic function of mode labels. Coefficients of a quadratic fit to these frequencies are denoted by $(\alpha, \beta, \gamma)$.
    {\bf b.} A stretched \'Echelle diagram where those five modes (square symbols) form a ridge-like structure.
    {\bf c.} Pulsation profile of TIC 422043310, which oscillates over a wide frequency range between 10 ${\rm d^{-1}}$ and 90 ${\rm d^{-1}}$. This star has been reported as a binary system, which is why it shows two distinct groups based on amplitudes. The quadratic spacing feature is seen in the high-amplitude group, suggesting that it is coming from the primary star.
    {\bf d.} Stretched \'Echelle diagram of TIC 422043310.
    {\bf e.} Multiple series of quadratic patterns seen in the power spectrum of TIC 138345546.
    {\bf f}. Various ridges in the stretched \'Echelle diagram, corresponding to the different series of quadratic patterns in the power spectrum of this star.}
    \label{qd-exmpl}
\end{figure*}

\begin{figure*}
    \centering
    \includegraphics[width=\textwidth]{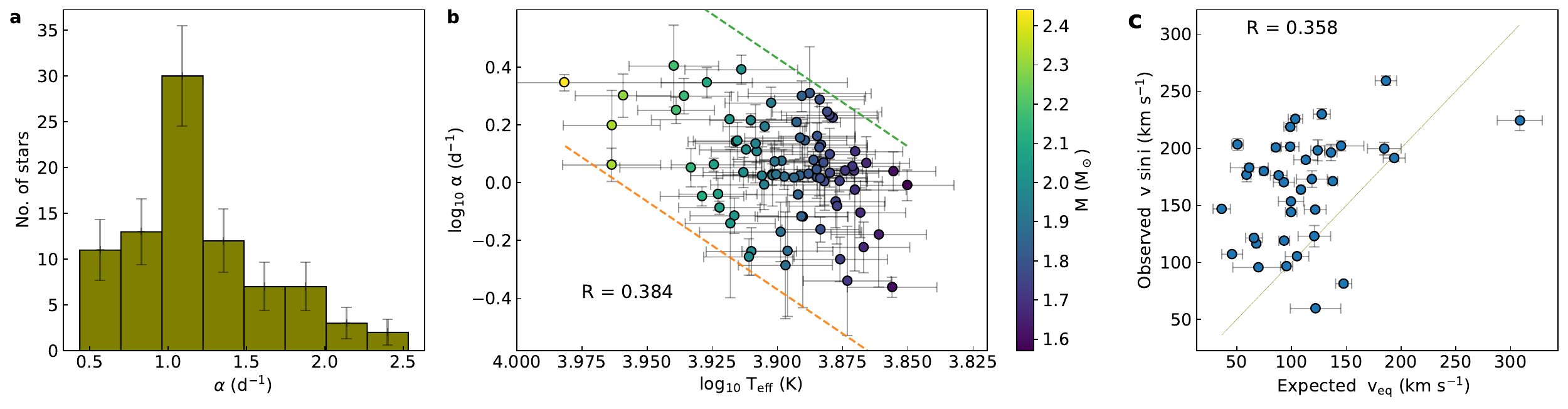}
    \caption{{\bf Coefficient $\alpha$ is related to stellar rotation.}
    {\bf a.} The distribution of $\alpha$ for 80 stars, seen to peak around $1 \rm d^{-1}$, is comparable to typical rotation rates of $\delta$ Sct stars.
    {\bf b.} Inferred rotational rates ($\alpha$) appear to be related to $T_{\rm eff}$. Points are color coded by stellar mass -- darker colors indicate lower masses.
    {\bf c.} Spectroscopically observed line-of-sight projected rotational velocities ($v \sin i$) plotted against $v_{\rm eq} = 2\pi \alpha R_\star$, i.e., treating $\alpha$ as if it were the equatorial rotation rate. The straight line marks $v \sin i = v_{\rm eq}$. Although we expect $v \sin i$ to be smaller than $v_{\rm eq}$, the similarity between the two independent inferences is suggestive of $\alpha$ being closely related to rotation. If our interpretation were to hold, $\alpha$ is systematically underestimated based on this plot.
    }
    \label{prf-a-rot}
\end{figure*}

Using a machine-learning based technique (see appendix \ref{Method}), we searched for the possible occurrences of unusual frequency arrangements in the oscillation spectra of 6711 $\delta$ Sct stars observed in the first 63 sectors of the TESS mission. We specifically studied the case of quadratic spacing, where mode frequencies are described as a quadratic function of mode labels. Thus, the frequency spacing between consecutive labels ($\Delta\nu$) does not remain constant, rather steadily increasing {{(e.g., $\Delta\nu_0, ~\Delta\nu_0+\delta\nu, ~\Delta\nu_0+2\delta\nu, ~\Delta\nu_0+3\delta\nu, ~\dots$ and so on)}}.

In 80 out of 6711 $\delta$ Sct stars, we found occurrences of 4 - 6 high-amplitude  oscillation modes, whose frequencies are arranged nearly quadratically. Equation \ref{e1} describes the sequence of resonant frequencies, where integer $i$ is an arbitrary  mode index (not necessarily radial order $n$ or azimuthal order $m$) and where $(\alpha, \beta, \gamma)$ are fitting constants, which we later demonstrate likely contain information pertaining to rotation,
\begin{equation} \label{e1}
    f_i = \alpha ~i^2 + \beta ~i + \gamma.
\end{equation}

Throughout our sample, TIC 305274761 is the only star found to be oscillating in five distinguishable frequencies, all of which are well described by a quadratic function of mode label (Fig. \ref{qd-exmpl}a). We also constructed an \'Echelle diagram (Fig. \ref{qd-exmpl}b) by appropriately stretching the frequency scale (for details, see Method) such that the quadratic pattern is converted to a linear pattern in the stretched frequency space. The straight ridge in the transformed \'Echelle diagram amplifies the significance of the quadratic pattern in (the unstretched) frequency space.

TIC 422043310, another noteworthy star from our sample, is listed as a binary system at 100.00\% probability in Gaia DR3 (\citealt{gaiadr3}), with a renormalised unit weight error (RUWE) of 3.095. Stars of RUWE values larger than unity are regarded as non-single objects (\citealt{Evans}). This star's power spectrum (see Fig. \ref{qd-exmpl}c) shows two interspersed but distinct groups of modes -- one of high amplitudes and the other with relatively lower amplitudes. A discernible quadratic frequency pattern is evident in the higher amplitude group -- suggesting that one component of this binary system is a $\delta$ Sct star oscillating in quadratically dispersed frequencies.

We also identified many stars whose power spectra carry multiple sequences of quadratically dispersed modes. TIC 138345546 is one such, with its pulsation spectrum shown in Fig. \ref{qd-exmpl}e. The appearance of multiple ridges in the transformed \'Echelle diagram (Fig. \ref{qd-exmpl}f) visually indicates the existence of three groups of quadratically dispersed modes across the oscillation spectrum. These features are reminiscent of solar-like spectra, where $\ell=0,2$ and $\ell=1$ form two distinct ridges.

Although this feature shares similarity with the multiple ridges seen in the \'Echelle diagrams of regularly oscillating $\delta$ Sct (\citealt{Nature2020}) and solar-like stars (\citealt{echlms}), the exact nature of modes forming these ridges are not so clear. We speculate each ridge may be consisting of modes of identical degree $\ell$ but different $m$. It should be noted that $m$ is no longer constrained to assume values between $-\ell$ and $\ell$, as the spherical-harmonic basis does not apply to descriptions of deformed spheres, such as in the case of rapidly rotating stars. Another possibility is that each ridge may be associated with non-asymptotic modes of differing radial order $n$ but identical $\ell$.

\subsection{Interpretation}

We obtained the coefficients ($\alpha, \beta, \gamma$) for each of the 80 stars by fitting equation \ref{e1} to their quadratically spaced resonances starting from $i=1$. The fit to $\alpha$ is invariant to choice of mode indices as long as they form a continuous sequence. For instance, if we fit the frequencies with a different set of mode indices ${i'}$ starting with $1+j$ instead of 1 - $j$ being an integer - leading coefficient $\alpha$ still retains its identical value, as explained in equation \ref{e2}. This indicates that $\alpha$ represents a physically important timescale. Although the unavailability of exact mode labels prevents us from exactly constraining $\beta$ and $\gamma$, the measurement of $\alpha$ does not face such issues.
\begin{align} \label{e2}
    f_i &= \alpha ~i^2 + \beta ~i + \gamma \nonumber \\
        &= \alpha (i'-j)^2 + \beta (i'-j) + \gamma \nonumber \\
        &= \alpha ~i'^2 +  (\beta -2 \alpha j) ~i' + (\alpha j^2 - \beta j + \gamma)
\end{align}

The distribution of parameter $\alpha$, shown in Fig. \ref{prf-a-rot}(a), peaks at around a unit cycle per day. Rotation rates of $\delta$ Scuti stars, calculated from their rotational velocities (\citealt{rot-ct}), are comparable to this in magnitude. Hence $\alpha$ may be related to rotation, although that is not the only explanation. For example, a stellar oscillation calculation obtained from GYRE (\citealt{GYRE_code}) for a model star of mass $1.7M_\odot$ evolved using Modules for Experiments in Stellar Astrophysics (MESA, \citealt{Paxton2019}), produced a small separation $\delta\nu$ (i.e. $\nu_{\rm n, \ell=0} - \nu_{\rm n-1, \ell=2}$) to be in range of $0.6-1.4 ~{\rm d^{-1}}$. Because the small separation is typically closely associated with linearly spaced harmonic degrees $\ell=0,2$, this explanation is unlikely to hold for the set of modes that we observe here. Regardless, additional observational evidence is crucial when investigating the origin and physics of these modes.

We obtained masses and effective temperatures of the 80 stars from the TESS Input Catalogue, whenever available. Fig. \ref{prf-a-rot}(b) shows the presence of a correlation between $\alpha$ and $T_{\rm eff}$, where points are color coded according to stellar mass. This trend is also consistent with the common intuition that cooler, low mass stars rotate more slowly than hotter heavier stars. The correlation between the two quantities, measured by the Pearson R coefficient, is 0.384.

Treating $\alpha$ as the equatorial stellar rotation rate and using radius values from the TESS Input Catalogue, we define ($v_{\rm eq} = 2\pi\alpha R_\star$) for these stars. We test this hypothesis by comparing with the spectroscopically measured (\citealt{gaiadr3}) line-of-sight projected rotational velocity ($v \sin i$), where $i$ is the inclination angle of the star's rotation axis. In Fig. \ref{prf-a-rot}(c), we compare the expected equatorial velocities with the observed $v \sin i$ counterparts for stars for which these measurements are available. This figure shows the existence of a connection between $\alpha$ and rotation in addition to a systematic offset. Pearson coefficient $R$ between the two measurements comes out to be 0.358, which although not large, suggests that the correlation between the two quantities is not altogether spurious.

The uniform frequency spacings that are seen in stellar pulsation spectra may be viewed as the $\alpha \rightarrow 0$ limit of a quadratically dispersed mode pattern. The examples we presented here represent frequencies that are non-linear functions of the mode indices. To assess how much better a quadratic polynomial fits these modes, we calculated their goodness-of-fit using the metric
$$\chi^2_{\rm reduced} = \dfrac{1}{N-3} \sum_{i=1}^N \dfrac{(\nu_i - \alpha i^2 -\beta i - \gamma)^2}{\sigma_\nu^2},$$
where $N$ denotes the number of modes in the identified sequence, $\nu_i$ their frequencies, $\sigma_\nu$ the uncertainty in frequencies ($0.2 {\rm d^{-1}}$) and the factor $N-3$ corresponds to the degrees of freedom given the quadratic polynomial is described with three coefficients. We obtained $\chi^2_{\rm reduced} < 1$ for 43 stars, implying that quadratic polynomials are adequate to fit the frequencies. The maximum $\chi^2_{\rm reduced}$ for our sample is 17. For the cases well fit using a quadratic relationship, the term $\alpha i^2$ is either comparable to or larger than the linear term $\beta i$, which is why a linear fit does not work. However, for stars whose modes are spaced apart at regular intervals, $\alpha\, i^2 << |\beta|\, i$ over the range of indices considered in the fit, in which event, a linear fit would produce $\chi^2_{\rm reduced}<1$.

\section{Discussion}

It may be tempting to presume that the marked modes in Figure \ref{qd-exmpl}(a) are 5 rotationally split multiplets corresponding to the sequence $m\in[-2,2]$. Should this hold true, the rotation rate would be set by the following equation using the multiplet frequencies of equal $|m|$ (\citealt{Deupree2011}).
\begin{equation*}
    f_{\rm rot} = \dfrac{f_{|m|}-f_{-|m|}}{2|m|}
\end{equation*}
Assigning indices $\rm m=\{-2, -1, 0, +1, +2\}$ to the modes in increasing order of frequency, the above equation yields $f_{\rm rot}\sim4 \rm ~d^{-1}$, which is greater than the Keplerian breakup rate for $\delta$ Sct stars (\citealt{rot-ct}). Additionally, the appearance of quadrupole modes without radial or dipole modes is not easily understood, since radial modes typically have greater visibility compared to non-radial modes. The second example (Figure \ref{qd-exmpl}c) presents another instance with which to rule out the rotational-multiplet description. We observed 6 modes, and assuming their $m$ values range from $-2$ to $+3$, the calculated rotation rate exceeds $10 \rm ~d^{-1}$, which is far beyond the Keplerian breakup rate.

Computational models of oscillations in rapidly oscillating stars have identified as 2-period island, 6-period island, whispering gallery and chaotic modes (\citealt{Lignieres2009}). All of these are described by dispersion relations that  are markedly different from the quadratic sequences we have identified.
We attempted through analogy to identify classical instances that resemble the dispersion relation presented here, i.e., $\omega\propto k^2$, where $k$ is spatial wavenumber and $\omega$ is temporal frequency. Standing waves on strings show $\omega \propto k$, small water ripples propagate as $\omega \propto k^{3/2}$, deep water waves as $\omega \propto \sqrt{k}$, shallow water waves as $\omega \propto k$, and light waves in plasma $\omega\propto \sqrt{\omega_p^2 + c^2k^2}$, where plasma frequency $\omega_p = 4\pi n_0 e^2/m$ defines the lowest frequency at which waves propagate through the plasma with electron density $n_0$. Although only tangentially relevant to the present analysis, fluid dynamics simulations of acoustic waves in a rapidly rotating ellipsoid (\citealt{quadratic_simulation}) show a quadratic dependence of mode frequencies with the rotational Mach number ($M_\Omega = a \Omega / c_s$), where $a$ represents the semi-major axis of the object and $c_s$ denotes the velocity of sound. Boundary conditions in each of these cases can serve to quantize the dispersion relation. At present, we are unable to find a suitable match to the observations, necessitating the future development of new theory and computation.

Determining the connection of index $i$ to the mode quantum numbers $(n,\ell,m)$ can help in identifying the origin of these unusual pulsations. Using MESA and GYRE, we performed several stellar evolution and oscillation calculations to support or rule out rotation as a cause of the quadratic frequency spacing. However, we found that regardless of whether we included rotation in GYRE or not, the frequencies of modes of a given harmonic degree $\ell$ always present as linear functions of the radial order $n$, as shown in the figure \ref{fig:GYRE}. We also tested several non-standard combinations of metallicity, Helium content, mixing lengths, convective overshooting etc. to test if any anomalous characteristics could explain the pulsations of these stars. In all cases, the radial mode frequencies primarily vary linearly with radial order. We include these details in Appendix \ref{sec:grid}.

Unlike the low-frequency Rossby modes (\citealt{rmode}), rotation does not serve as a restoring mechanism for the modes we have discussed here, since the time-scales of these acoustic high-frequency oscillations are much shorter than that associated with rotation. Nevertheless, rotation perturbs the mode frequencies, thereby modifying their organization in the pulsation spectra.

\appendix 

\section{Methods}

Searching for new features among the large sample of stars observed by TESS through manual visual inspection is infeasible. We thus devised a neural-network based automated method to accelerate this search routine. We trained the network with $\sim1$ million $\delta$ Sct-like spectra, constructed in an ad-hoc manner, which we describe below. Our approach is motivated by the artificial model-generation strategy developed for solar-like pulsators (\citealt{Benomar2008}) to enable inferences, though significant adjustments were required to adapt the technique to $\delta$-Scuti-like pulsations.

\subsection{Simulating synthetic spectra}

Each spectrum was constructed with 5 to 6 modes whose frequencies were set to be quadratic functions (equation \ref{e1}) of mode labels with different combinations of $(\alpha, \beta, \gamma)$, sampled between the range $[0.1-3.0]$, $[0.1-6.0]$ and $[4.0-10.0]$ respectively, all in the units of ${\rm d}^{-1}$. These parameter ranges were set by ensuring that the simulated frequencies remain well within the usual range exhibited by $\delta$ Scutis. We additionally incorporated uniformly spaced frequencies (\citealt{Reese2006, PhysRevLett.107.121101}) and added peaks at arbitrary locations representative of chaotic modes (\citealt{Lignieres2009}). These modes were assigned  arbitrary amplitudes between 0 and 1 sampled from a log-uniform distribution. Because these stars are observed over temporal windows much shorter than mode lifetimes, we convolved each mode with a sinc function (${\rm sinc} (\pi \nu / 2f_{\rm Nyquist})$, \citealt{revPL2022}). Finally, we superimposed a background noise by adding absolute values of amplitudes sampled from a Gaussian distribution ($\mu=0, ~\sigma=1$) and scaled arbitrarily between $10^{-3}$ and $10^{-1}$. An example of the simulated spectra is shown in Figure \ref{fig:evl-cnn}. Our aim is not to generate altogether realistic spectra, given the limited understanding of $\delta$ Scuti oscillations -- rather, we intended to train a machine-learning model on pulsation spectra to identify quadratic dispersions and map them to the relevant coefficients.

\subsection{Machine learning model} \label{Method}

\begin{figure*} 
  \centering
  \includegraphics[width = \textwidth]{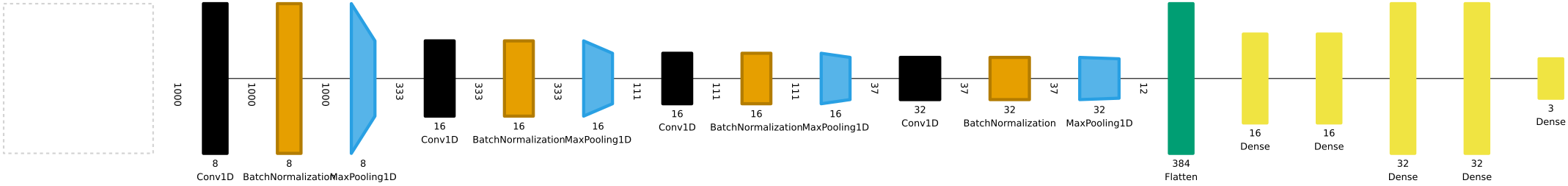}
  \caption{Network architecture used in the analysis.}
  \label{fig:cnn-architecture}
\end{figure*}

To predict values of $(\alpha, \beta, \gamma)$, a 1D deep convolution network was designed with \textsc{Tensorflow} (\citealt{tensorflow}) and \textsc{Keras} Python packages and was trained on the simulated spectra. The input is a 1000$\times$1 dimension vector representing the simulated power spectrum. The network comprises convolution layers followed by fully connected layers and the related parameters are tabulated in Table \ref{cnn-parameter}, with the architecture demonstrated in Figure \ref{fig:cnn-architecture}. This architecture and its hyperparameters were chosen following the standard designs commonly found in a wide range of CNN models (\citealt{FERREIRA2018205}). Such networks are generally referred to as convolutional neural networks (\citealt{convolutional}) which are efficient in visualization and pattern extraction from images. It suits our purpose to identify quadratic sequences from thousands of stellar spectra. The network was trained using mean-squared error loss and the Adam optimizer (\citealt{adam}). To ensure good convergence of the model, a learning rate schedule (\citealt{decaylearningrate}) was employed to reduce the learning rate after each epoch and we trained the network for 170 epochs, at which point the loss function plateaued.

\begin{table} 
\centering
\caption{Hyperparameter configuration used in our network.}
\begin{tabular}{ll}
    \hline \hline
    Hyperparameters & Values \\ \hline
   Convolution Layers & 4 \\
   \qquad kernel size & (5x1)  \\
   \qquad No. of filters & 8, 16, 16, 32  \\
   Fully connected layers & 5 \\
   \qquad layer size & 16, 16, 32, 32, 3 \\
   Loss type & mean squared error \\
   Learning rate & 4e-3, 1e-3\\
   Batch size & 64 \\
  \hline
\end{tabular}
\label{cnn-parameter}
\end{table}

\begin{figure*}
    \centering
    \includegraphics[width=1\linewidth]{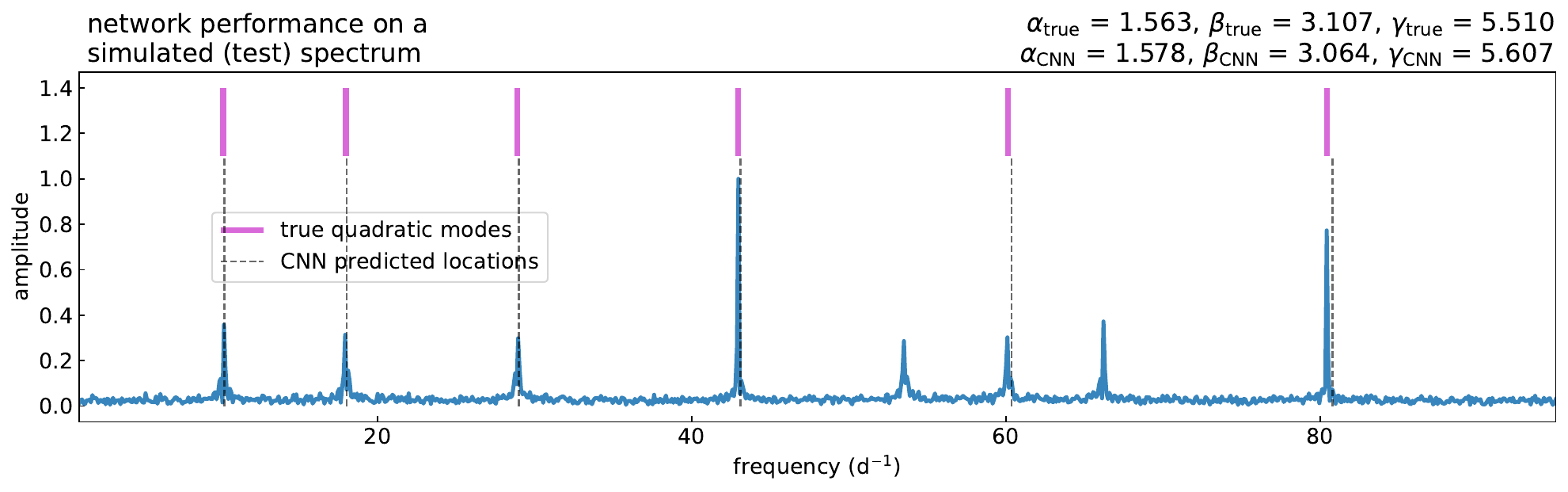}
    \caption{Demonstrating the method's performance on a simulated (test)  $\delta$-Scuti-like oscillation spectrum with 6 quadratically arranged frequencies and 2 chaotic modes. The original ($\alpha, \beta, \gamma$) values used to construct this spectrum are specified in the figure title, along with the corresponding values inferred by the CNN. Based on this inference, we mark with dashed lines the expected frequency location where the quadratic modes may appear. Also shown are the positions of the original quadratic modes in the simulated spectrum, to which CNN's identifications are close.
    }
    \label{fig:evl-cnn}
\end{figure*}

We trained the network over 75\% of the simulated spectra leaving the remaining 25\% as validation data for subsequent evaluation of the CNN over which it was able to infer the $\alpha$, $\beta$, and $\gamma$ values with a mean combined error of 3.4\%. Since the test data was unseen by the network, performance over this sample reflects the extent to which the CNN has generalized. In Figure \ref{fig:evl-cnn}, we show one of the test spectra over which the trained CNN was applied to infer ($\alpha, \beta, \gamma$) values. Inferred values are comparable to the original values used to construct the spectrum, with 1\%, 1.3\%, and 1.7\% relative error in $\alpha$, $\beta$ and $\gamma$ respectively. Corresponding modes emerging from this inference are also in reasonable agreement with the simulated frequencies, which demonstrates the network's efficiency in discerning quadratic patterns from power spectra.

\subsection{Identifying potential stars}

We applied the trained neural network model over the 6711 TESS $\delta$ Scuti stars catalogued in Singh et al. (2024). Power spectra of these stars were introduced to the network as inputs to generate the corresponding $(\alpha, \beta, \gamma)$ values. These coefficients were used to mark the expected mode locations in the power spectra of these stars. Notably, in 80 stars, we found the presence of significant peaks in close proximity of the expected mode locations. Finally, we extracted the exact frequencies of these observed peaks and fit these frequencies with a quadratic polynomial to obtain the $(\alpha, \beta, \gamma)$ coefficients.

\subsection{Stretched \'Echelle diagram}

The \'Echelle diagram serves as a potent tool for clustering periodic patterns present in spectra. It is constructed by fragmenting the entire spectrum into equal-width segments and stacking them vertically. In solar-like spectra for instance, modes of identical harmonic degree ($\ell$) but varying radial order ($n$) form vertically aligned ridge-like structures in this diagram. However since our spectrum contains non-linear features, ordinary \'Echelle diagrams cannot display meaningful structure. Following (\citealt{Mosser}) we develop a modified-\'Echelle diagram wherein we stretch the frequency domain so as to transform the non-linear pattern into a linear analogue. Straightforward algebraic manipulation in Equation \ref{e1} gives us
\begin{align}
    f_i &= \alpha \left(i+\dfrac{\beta}{2\alpha}\right)^2 + \left(\gamma-\dfrac{\beta^2}{4\alpha}\right),
\end{align}
which may be further modified (equation \ref{e6}) to linearize the mode index $i$,
\begin{align} \label{e6}
    \implies \Tilde{f} = \sqrt{f+\left(\dfrac{\beta^2}{4\alpha}-\gamma \right)} = \sqrt{\alpha} \left(i+\dfrac{\beta}{2\alpha}\right).
\end{align}
This resembles the asymptotic formula $\nu = \Delta\nu (n+0.5\ell+\epsilon)$ which determines the frequencies of pressure-mode oscillations in many stars. It allows us to treat the $\Tilde{f}$ space as though in the frequency domain, and $\sqrt{\alpha}$ as the width with which the $\Tilde{f}$ space is to be folded (followed by vertical stacking) in order to construct the {transformed \'Echelle diagrams}.

\subsection{Computing stellar models} \label{sec:grid}

We constructed a grid of 100 stellar models by varying the mass ($M$), initial Helium ($Y_i$) and metal ($Z_i$) content, overshooting parameter ($f_{\rm OV}$), mixing length parameter ($\alpha_{\rm MLT}$), and age ($\tau$) over a wide range given in Table \ref{grid_range}. Following the method described by \cite{grid_resource}, we evolved non-rotating stellar models using \texttt{MESA} - the calculations incorporated the Eddington $T-\tau$ atmosphere, mixing length theory of convection, and exponentially decaying overshooting across the convective boundaries. We assigned the stellar models arbitrary (uniform) rotation rates between $0-2.0 {\rm d^{-1}}$. The stellar oscillation code \texttt{GYRE} was used to compute frequencies of acoustic modes of degree $\ell=0$ and radial orders between $1-9$, while implementing the traditional approximation of rotation. Investigating solutions through the grid, we observed that all model pulsations exhibit p-mode frequencies that are linear functions of radial orders. Thus, we are unable to find stellar models that can potentially explain the quadratic frequency dispersion.

\begin{table}
    \centering
    \begin{tabular}{cccccc}
    \hline \hline
        $M (M_\odot)$ & $Y_i$ & $Z_i$ & $\alpha_{\rm MLT}$  & $f_{\rm OV}$ & $\tau$ (Myr) \\ \hline
         1.5 & 0.2 & 0.004 & 0.5 & 0.002 & 1 \\
         2.5 & 0.4 & 0.050 & 3.0 & 0.060 & 1000 \\ \hline
    \end{tabular}
    \caption{Ranges of parameters within which the stellar models grid was developed.}
    \label{grid_range}
\end{table}

\begin{figure}
    \centering
    \includegraphics[width=0.49\linewidth]{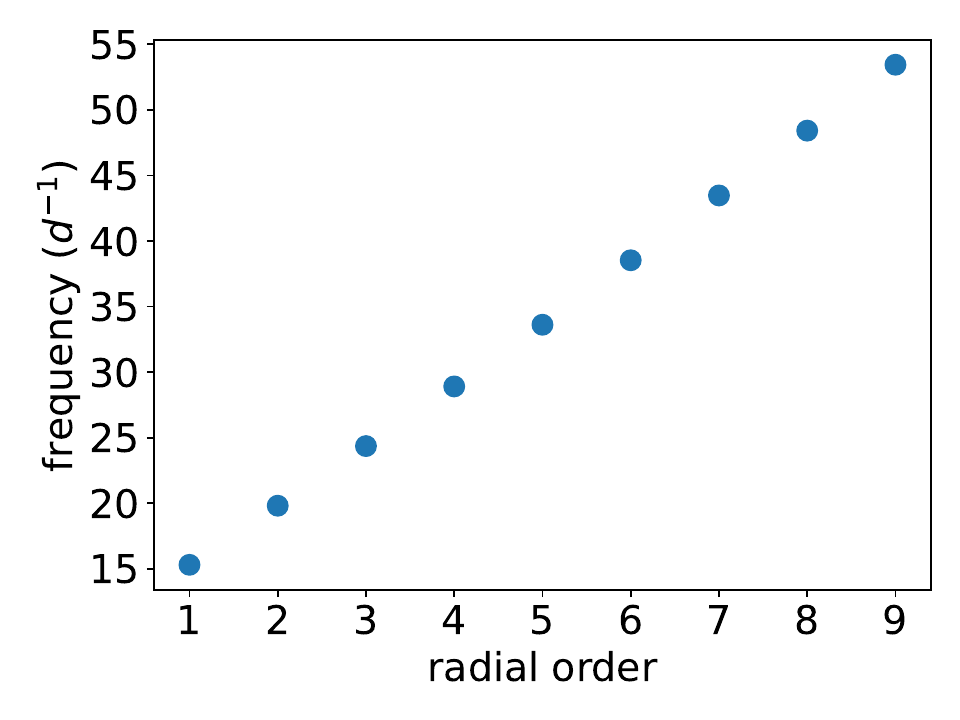}
    \caption{Frequencies of $\ell=0$ modes as a function of radial order $n$ for a $\delta$ Scuti-like star of $M=1.9M_\odot$, $Y_i=0.30$, $Z_i=0.020$ at the age of 526.4 Myr, computed using \texttt{MESA} and \texttt{GYRE}. We implemented a uniform rotation rate of $1.0 {\rm d^{-1}}$ using the traditional approximation available in \texttt{GYRE}. No quadratic frequency dispersion was observed, even in the non-asymptotic regime.}
    \label{fig:GYRE}
\end{figure}

Acknowledgments: We acknowledge the useful insights we received from Dr. Daniel R. Reese, Dr. Fran\c{c}ois Ligni\`eres, and Dr. Timothy R. Bedding. This work has made use of the SIMBAD and VizieR databases. We are thankful for the valuable data released by the NASA's TESS mission and the European Space Agency (ESA) space mission Gaia. We applied \texttt{Lightkurve} (\citealt{Lightkurve}), a Python package for treating Kepler and TESS data. We  have performed all computations in the Intel Lab Academic Compute Environment. We also acknowledge support from the Department of Atomic Energy, Government of India (grant RTI 4002). This research was supported in part by a generous donation (from the Murty Trust) aimed at enabling advances in astrophysics through the use of machine learning. Murty Trust, an initiative of the Murty Foundation, is a not-for-profit organisation dedicated to the preservation and celebration of culture, science, and knowledge systems born out of India. The Murty Trust is headed by Mrs. Sudha Murty and Mr. Rohan Murty. We are grateful to the reviewer for their useful and constructive comments.

{\it Supplementary Resources:}
A text file is available with the TIC IDs of the stars from our dataset, their ($\alpha$, $\beta$, $\gamma$) coefficients, observationally determined ($T_{\rm eff}$, $v \sin i$) values, and  their $\chi^2_{\rm reduced}$ metrics. Additionally, a pdf file is provided showing the oscillation spectra of these stars, highlighting their quadratic frequency patterns, and demonstrating their stretched \'Echelle diagrams. We also provide the MESA and GYRE inlist used for this research.

\end{document}